\documentclass[
 reprint, bibnotes,
 amsmath,amssymb,
 aps, prl
]{revtex4-1}
\usepackage{graphics,epsfig}
\usepackage{color}
\usepackage{graphicx}
\usepackage{dcolumn}
\usepackage{bm}
\usepackage{hyperref}
\usepackage{enumitem}
\usepackage{placeins}

\begin{document}
\title{Diversity, stability, and reproducibility in stochastically assembled  microbial ecosystems}
\author{Akshit Goyal}
 \affiliation{The Simons Centre for the Study of Living Machines, NCBS-TIFR, Bengaluru 560 065, India.}
\author{Sergei Maslov}%
\thanks{maslov@illinois.edu}
\affiliation{%
 Department of Bioengineering and Carl R. Woese Institute for Genomic Biology,
University of Illinois at Urbana-Champaign, Urbana, IL 61801, USA.}

\date{\today}

%--------------------------------------------------------------------------------------------------
% Abstract
%--------------------------------------------------------------------------------------------------
\begin{abstract}
Microbial ecosystems are remarkably diverse, stable, and often consist of a balanced mixture of core and peripheral species. Here we propose a conceptual model exhibiting all these emergent properties in quantitative agreement with real ecosystem data, specifically species’ abundance and prevalence distributions. Resource competition and metabolic commensalism drive stochastic ecosystem assembly in our model. We demonstrate that even when supplied with just one resource, ecosystems can exhibit high diversity, increasing stability, and partial reproducibility between samples.
\end{abstract}

\maketitle

%--------------------------------------------------------------------------------------------------
% Introduction section
%--------------------------------------------------------------------------------------------------
% \section*{Introduction}
Natural microbial ecosystems are remarkably diverse, often harboring hundreds to thousands 
of coexisting species in microscopic volumes \cite{Lozupone2012, Curtis2002, Mei2016}. 
How do these ecosystems manage to acquire and maintain such a high diversity? 
This so-called `paradox of the plankton' \cite{Hutchinson1961} is especially surprising given that 
microbes are capable of rapid exponential growth and fierce competition for nutrients. 
Indeed, the competitive exclusion principle \cite{Curtis2002, Hsu1977} postulates 
that the number of species in an ecosystem at steady 
cannot exceed the number of available nutrients.

Compounding this puzzle, theoretical studies \cite{May1972} suggest 
that highly diverse ecosystems are generally prone to instabilities. 
This brings up a second question: how do naturally-occurring 
microbial ecosystems manage to remain relatively stable despite their diversity?

Moreover, ecosystems operating under similar environmental conditions could be rather different from each other in terms of species composition \cite{Caporaso2011, Mei2016, Barberan2012}. 
This apparent lack of reproducibility does not apply equally to different organisms. 
Some species, classified as `core' or `keystone', are detected in most individual 
ecosystems. Other `peripheral' species are only observed in a small fraction of 
them. Observed species' prevalence distributions (the fraction of similar ecosystems a species 
is detected in) are often U-shaped: their peaks occupied by these core and peripheral species respectively. 
We are thus presented with a third question: what determines the reproducibility (or lack thereof) 
of species composition in microbial ecosystems? 
%Understanding reproducibility (or lack thereof) of the species 
%composition of microbial communities 
%causes and determinants of 
%species composition and prevalence 
%what causes such patterns 
%of species composition 
%is the third major focus of research in microbial ecology.

Here, we introduce a conceptual model of a stochastically assembling 
microbial ecosystem, which in spite of its simplicity, addresses and 
suggests possible solutions to all three of these long-standing puzzles.

To explain the aforementioned high diversity and poor reproducibility, previous models have relied on a number of factors including spatial heterogeneity \cite{Hsu1977, Huisman1999}, temporal and seasonal variations in resource availability \cite{Pfeiffer2001, MacLean2006}, thermodynamic constraints \cite{Grosskopf2016}, microbial `warfare' and cooperation via ecological feedbacks \cite{Czaran2002}, and predation by bacteriophages \cite{Sneppen2015, Maslov2017}.
In contrast to this, our model attributes high diversity to metabolic byproducts secreted 
by microbes due to incomplete resource-to-biomass conversion, which could in turn be used by other species for growth. 
By its very nature, our model simultaneously exhibits (a) high species diversity, (b) 
gradually increasing stability, reached after repeated rearrangements, (c) a U-shaped 
prevalence distribution and (d) a positive abundance-prevalence correlation.

%--------------------------------------------------------------------------------------------------
% Figure 1
%--------------------------------------------------------------------------------------------------
\begin{figure*}[htbp]
% figure 1
\includegraphics[width=\textwidth]{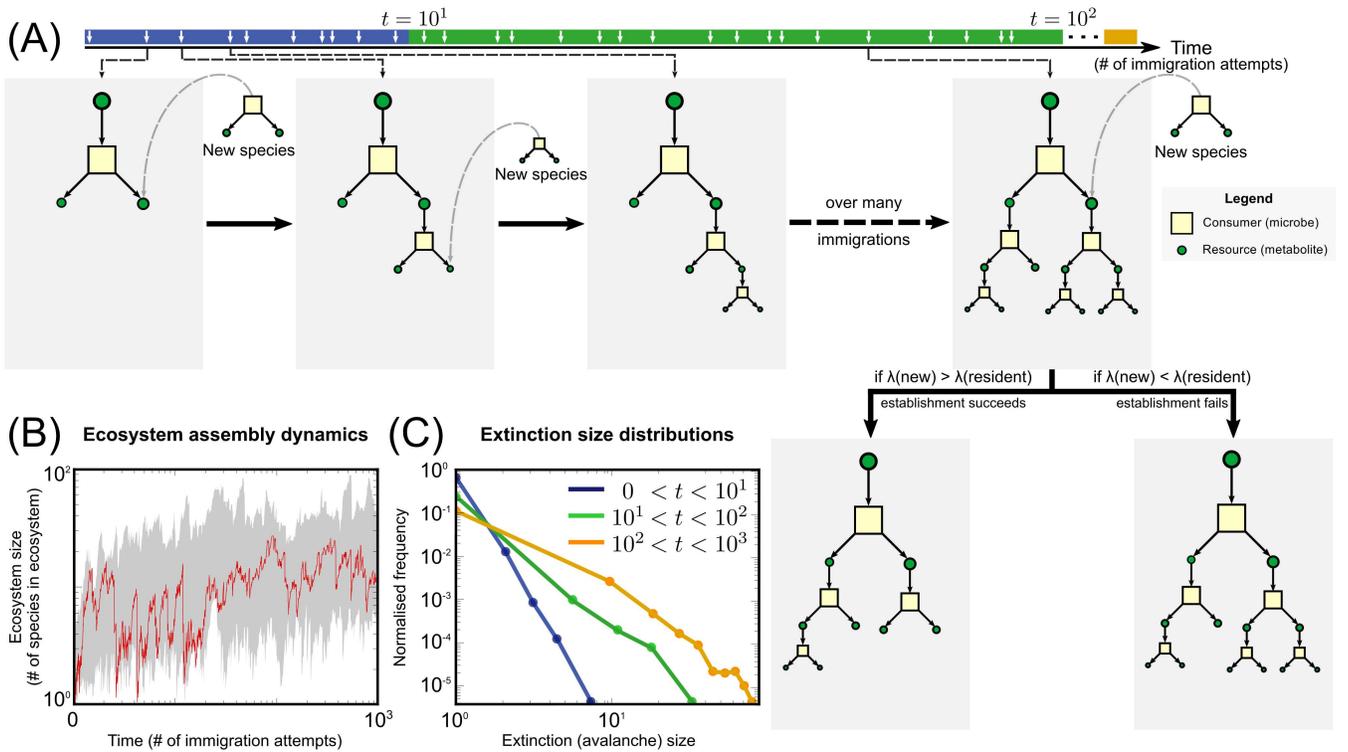}

\caption{\scriptsize{\textbf{Ecosystem assembly in the model.} (A) The diagram illustrates different phases in the assembly dynamics involving species (yellow squares) consuming resources (green circles). Sizes are indicative of the steady state abundances and concentrations. 
Initially, only a single externally supplied resource (the largest green circle) is available and consumed by a microbe, which in turn secretes $\beta = 2$ metabolic byproducts. New species immigrate into this ecosystem 
%at Poisson rate 1 
(immigration events marked on the timeline), each using only one resource. Ecosystem establishment is contingent on the following assembly rule: if the resource affinity $\lambda$ of the new species is higher than any resident species on its chosen resource, the immigrant species survives and the resident goes extinct (along with all its dependents). (B) A sample assembly trajectory (in red) of the ecosystem size (number of species) as a function of 
time ($t$, measured in number of 
immigration attempts) at dilution rate $\delta=10^{-1}$ days$^{-1}$.
%corresponding to a residence time ($\delta^{-1}$) of roughly 10 days. 
The gray envelope shows ecosystem sizes over 1,000 assembly trajectories. (C) Extinction size distributions (number of species that go extinct during a single immigration event) get broader as ecosystem assembly proceeds: $t < 10^1$ (blue); $10^1 < t < 10^2$ (green) and $10^2 < t < 10^3$ (orange).}}
\label{keyfig1}
\end{figure*}

While our model clearly does not include many of the previously proposed 
factors known to affect these features, 
we believe it is a reasonable first order description of some real ecosystems,  
examples of which include the human oral microbiome \cite{Caporaso2011}, 
methanogenic bioreactors \cite{Nobu2015}, and anaerobic digesters in wastewater treatment 
plants \cite{Mei2016}.

%--------------------------------------------------------------------------------------------------
% Results section
%--------------------------------------------------------------------------------------------------
% \section*{Results}
% \subsection{General outline of the model}
% \vskip 5pt
Our model describes a dynamic microbial ecosystem in which species attempt to populate the environment externally supplied with a single resource. 
We assume that species can convert only a fraction of consumed resources 
into their biomass, while secreting the rest as metabolic byproducts. 
%These byproducts 
These in turn may serve as nutrient sources for other species in the ecosystem. 
This allows even one externally supplied resource to support high ecosystem diversity purely 
via byproduct-driven commensal interactions. 

% The ecosystem co-inhabits a bioreactor-like environment diluted at a fixed rate. 
%
% To simplify our mathematical analyses, we assume that each species can utilize just one resource. A more general case of multiple resources per species (yet utilized one at a time such as in diauxic growth) leads to rather similar dynamics. 

New species are constantly introduced to this environment from some external population. Their survival or extinction is determined by a simple rule dictated by competitive exclusion. Because of the commensal relationship between these species, elimination of just one species 
%due to its loss in the competition with a newly introduced 
may lead to an `extinction avalanche' in which multiple species are lost.

We explore how species diversity in microbial ecosystems is established over time. Moreover, by 
%running the colonization experiment multiple times, 
simulating several instances of ecosystem assembly, 
we can separate the set of core (high-prevalence) species from those with progressively lower prevalence. 
%For certain time periods, 

% \subsection{Steady state abundances and fluxes in a bioreactor} 
The dynamics in our bioreactor-like environment is fully characterized by the 
concentrations of individual resources (metabolites) labeled as $C_0, C_1, . . .$ and the abundances of all resident microbial species labeled as $B_1, B_2, . . .$.  When we initialize the model, each species is assigned a single resource it can grow on and $\beta = 2$ metabolic byproducts. All resources are randomly selected from a `universal chemistry' of size $N_{\text{univ}} = 5,000$. This choice is inspired by the total number of metabolites in KEGG's metabolic database \cite{Kanehisa2000}. However, qualitatively similar results are obtained for much smaller values $N_{\text{univ}}$, for example, the number of carbon sources typically utilized by microbes.

The environment is supplied with a single resource (labeled $0$) at a constant flux $\phi_0$. After several attempts, the first microbial species (labeled $1$) capable of utilizing the resource $0$ colonizes the environment. The following equations determine the dynamical behavior of resource concentration $C_0$ and microbial abundance $B_1$:

\begin{align}
% \frac{dC_0}{dt} = \phi_0 - \lambda_1 C_0 B_1 - \delta \cdot C_0,
% \end{equation}
%
% \begin{equation}\label{B1_eq}
% \frac{dB_1}{dt} = \lambda_1 C_0 (1 - \alpha) B_1 - \delta \cdot B_1,
\label{C0_eq}\frac{dC_0}{dt} &= \phi_0 - \frac{\lambda_1 C_0 B_1}{Y} - \delta \cdot C_0, \\
\label{B1_eq}\frac{dB_1}{dt} &= \lambda_1 C_0 B_1 - \delta \cdot B_1,
\end{align}

Here, $\lambda_1 C_0$ is the growth rate of the species $1$ 
consuming the resource $0$ at a rate $\frac{\lambda_1 C_0}{Y}$, where 
$Y$ is the yield of the biomass conversion process 
(the number of microbes per unit concentration of the resource). 
The resource affinity $\lambda$ is assigned by a random draw from a log-normal 
distribution such that the logarithm of $\lambda$ has mean 0 and variance 1. 
Note that using a more general expression for microbial growth as a function of nutrient 
concentration, e.g. Monod's law, does not affect our results. 

Our model is based on carefully following the flow of resources (e.g. carbon)
throughout the ecosystem. Different resources could be inter-converted into each 
other and into the biomass of microbes. Hence it is convenient to measure 
all microbial abundances in units of the 
resource concentration. 
%This will make the interconversion of 
%different resources easier to trace. 
We adopt this change of units for $B_i$ for the rest of this manuscript. 
Microbial yield is given by $Y=(1-\alpha)$ where $(1-\alpha) < 1$ represents the fraction of 
the consumed resource (e.g. carbon atoms) successfully converted to biomass. 
The remainder is secreted as two byproducts 
$1$ and $2$ getting shares $\nu_1 \alpha$ and $\nu_2 \alpha =(1-\nu_1) \alpha$ 
respectively. 

Another interpretation of these equations would apply if all processes 
in the ecosystem were energy-limited (as opposed to nutrient-limited). 
In this case, it would be convenient to measure the concentrations of 
both resources and microbes in units of the energy density. 
The factor $(1 - \alpha)$ could then be interpreted as an 
energy conversion efficiency. Due to dissipation, in this case it would be possible 
for $\alpha$ (the fraction of the incoming energy flux secreted as byproducts)
to be smaller than the leftovers from biomass conversion.
Barring small corrections, the results of our model would be equally 
applicable to such energy-limited ecosystems. 

% Species $1$ converts this resource to biomass with an efficiency $(1 - \alpha)$. 
% %Thus, the fraction of the initial resource flux converted to microbial biomass 
% %is $\phi_0 \cdot (1-\alpha)$. 
% We further assume that the remaining flux $\phi_0 \cdot \alpha $ 
% is secreted by the species in the form of two metabolic byproducts $1$ and $2$ getting  shares $\nu_1$ and $\nu_2=1-\nu_1$ of this flux. In the model, the number $\nu_1$ assigned to each species is independently drawn from a uniform distribution between $0$ and $1$.

We assume that the concentrations of both microbes and resources are diluted at the same rate, $\delta$. It is straightforward to generalize our model to a case where these dilution rates are in fact different (as is often the case in batch-fed bioreactors). Throughout this manuscript we are only interested in the steady state properties of the system, which can be easily derived from equations (\ref{C0_eq}) and (\ref{B1_eq}). At steady state, $C^*_0$ and $B^*_1$ are given by:

\begin{align}
\label{C0*_exp}C_0^* &= \frac{\delta}{\lambda_1},\\
\label{B1*_exp}B_1^* &= \frac{\big(\phi_0 - \frac{\delta^2}{\lambda_1}\big) Y}{\delta} = \frac{\widetilde{\phi_0} (1 -\alpha)}{\delta}.
\end{align}

Here, to simplify our notation, we have introduced the effective flux of a nutrient (adjusted for dilution) which is given by  $\widetilde{\phi_0} = \phi_0 - \delta \cdot C_0^* = \phi_0 - \frac{\delta^2}{\lambda_1}$. Note that (a) at steady state resource concentration $C_0^*$ depends inversely on $\lambda$ implying that if two species were to compete for the same resource, the one with a higher $\lambda$ would drive the resource concentration lower than the other, thus being the only survivor of the two, and (b) unlike the steady state nutrient concentration, the steady state species abundance is largely independent of $\lambda$. Indeed $\lambda$ only enters this equation via the effective resource flux which in the limit of low dilution approximates $\phi_0$.

\FloatBarrier

%--------------------------------------------------------------------------------------------------
% New species
%--------------------------------------------------------------------------------------------------
% \subsection{Stochastic assembly rules}

We simulate ecosystem assembly in discrete time steps corresponding to 
%. At each new step, we 
the introduction of a new microbial species into the ecosystem. 
We assume that these events are sufficiently infrequent for the system
to reach steady state between two subsequent immigration attempts. 
We measure time in the number of attempted species immigrations. As explained before, each species in our model 
uses just one resource and consumes it with a randomly drawn affinity $\lambda$. 
If the resource consumed by a newly introduced species is available, 
the outcome can be one of two possibilities: (a) if said resource is unused by any other species, the newcomer survives and becomes a resident, or (b) if said resource is used by a current resident, the species survives only if it has a larger $\lambda$ than the species currently using it.

If the newly introduced species survives, its abundance at steady state is determined by the same expression as in equation (\ref{B1*_exp}), replaced with the effective flux %$\widetilde{\phi_i}$ 
of the resource that it consumes. 
If all byproducts are equally partitioned, the average effective flux at trophic layer $\ell$ will be related to the 
external resource flux via:
\begin{equation}
\label{fluxEqn}
\langle \widetilde{\phi_i} \rangle_{\ell} = \phi_0 \Big(\frac{\alpha}{\beta}\Big)^\ell - \frac{\delta^2}{\lambda} \Big( 1 + \frac{\alpha}{\beta}\Big)^\ell.
\end{equation}

Note that if a new species survives by competitively displacing another, 
it could also lead to the extinction of any species that directly or indirectly 
depend on the latter for byproducts. As ecosystem assembly proceeds, we observe 
that the distribution of the extinction size (the number of species that go extinct 
during such an event) gets broader over time (see figure \ref{keyfig1}(C)). 

Over many steps of ecosystem assembly, as species use and secrete more byproducts, the entire ecosystem assumes a tree-like structure (see figure \ref{keyfig1}(A) for ecosystem structure and \ref{keyfig1}(B) for sample dynamical trajectories). 

%--------------------------------------------------------------------------------------------------
% Figure 2
%--------------------------------------------------------------------------------------------------
\begin{figure}[t]
% figure 2
\includegraphics[width=0.48\textwidth]{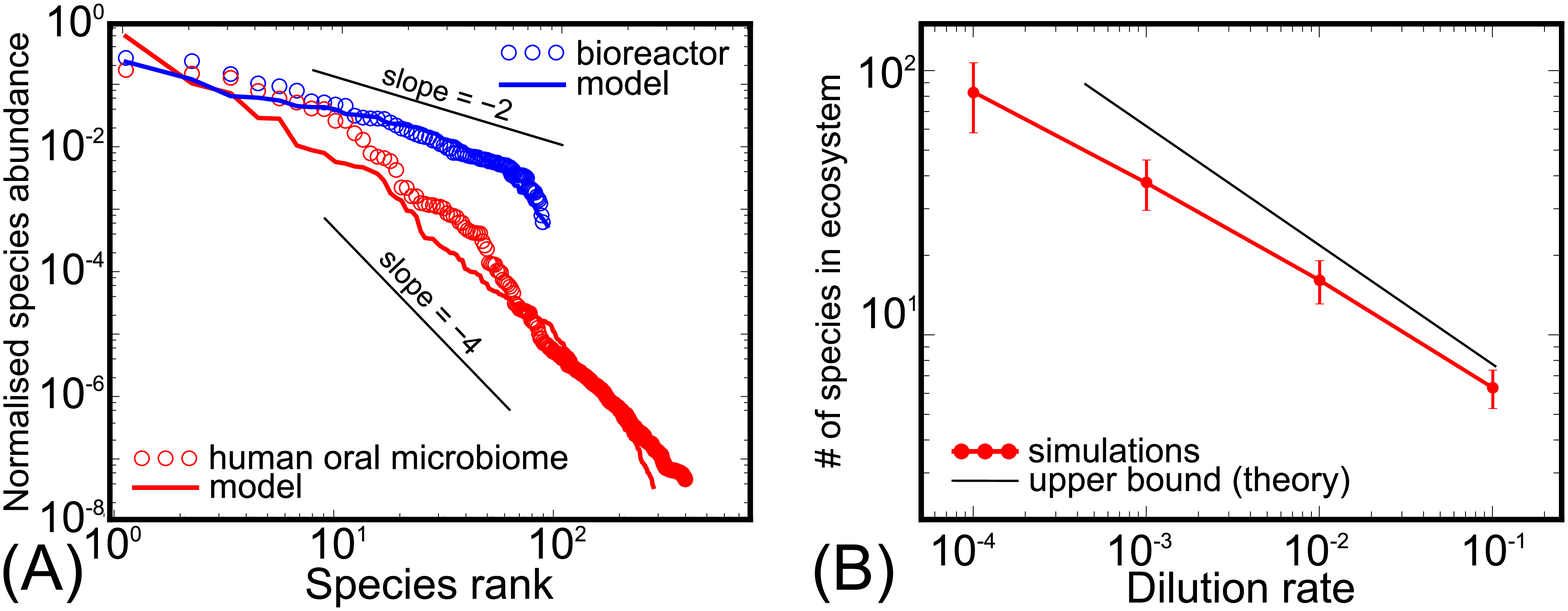}

\caption{\scriptsize{\textbf{Emergent ecological features.} (A) Rank-abundance plot of normalized species abundances in a methanogenic bioreactor \cite{Nobu2015} (blue circles) and the human oral microbiome \cite{Caporaso2011} (red circles) and for comparison, simulated ecosystems from our model (corresponding solid lines) with $\alpha$ equal to 0.5 and 0.1, respectively. (B) The dilution rate $\delta$ in the chemostat controls the maximal size $\mathcal{N}_\text{max}$ of the ecosystem coexisting on a single externally-supplied resource. Here, $\alpha=0.1$ and $\beta=2$. $\mathcal{N}$ approximately agrees with the expression in equation (\ref{dilScalingEqn}).}}
\label{keyfig2}
\end{figure}

% \subsection{Emergent ecological features}
% We thus demonstrate a possible assembly mechanism by which a microbial ecosystem may have more than one species coexist on a single (externally-supplied) resource through dependence on species' metabolic byproducts. Metabolic byproducts in this ecosystem create `niches' which new incoming species may occupy and survive.

Moreover, our model ecosystems have two interesting emergent features. Firstly, species' steady-state abundances follow a power-law. To see this, note that these abundances depend on the effective flux of the consumed resources. Typically, for a species at trophic layer $\ell$, $B^* \approx \frac{\langle \widetilde{\phi_i} \rangle_{\ell} (1 - \alpha)}{\delta}$. Now each layer can accommodate $\beta^\ell$ species, where $\beta$ is the number of byproducts per species. Using this, we can show that $\mathcal{N}(b)$, the number of species with abundance $b$ follows (see supplement \cite{SuppMat} for derivation):

\begin{equation}
    \mathcal{N}(B = b) \sim b^{-\big(1 + \frac{\text{log}\beta}{\lvert\text{log}\alpha\lvert + \text{log}(\alpha + \beta)}\big)}.
\end{equation}

In a rank-abundance plot (see figure \ref{keyfig2}(A)), species abundances follow a power-law with exponent $\frac{\lvert\text{log}\alpha\lvert + \text{log}(\alpha + \beta)}{\text{log}\beta}$. For appropriately chosen $\alpha$, both this expression and our simulations (solid curves in figure \ref{keyfig2}(A)) agree with data from microbial ecosystems sampled from the human tongue \cite{Caporaso2011} and methanogenic bioreactors \cite{Nobu2015} (open circles in figure \ref{keyfig2}(A)).

Secondly, the dilution rate $\delta$ sets a limit to the number of species in the ecosystem. This happens when the resource flux (as in equation (\ref{fluxEqn})) at the bottom-most layer $\ell_\text{max}$ becomes negative. We can show that the number of species in the ecosystem $\mathcal{N}_\text{max}$ (proportional to $\beta^{\ell_{\text{max}}}$) follows (see supplement \cite{SuppMat} for derivation):

\begin{equation}
    \label{dilScalingEqn}
    \mathcal{N}_\text{max}(\delta) \sim \delta^{-\big(\frac{2\text{log}\beta}{\lvert\text{log}\alpha\lvert + \text{log}(\alpha + \beta) } \big)} .
\end{equation}
For $\beta = 2$ and $\alpha = 0.1$, this expression (black solid line in figure \ref{keyfig2}(B)) approximates our simulated ecosystems (red solid line in figure \ref{keyfig2}(B)). Note that this expression provides an upper bound to the number of species. It assumes both equal partitioning of all byproducts and equal $\lambda$s for all species.

%--------------------------------------------------------------------------------------------------
% Repeated assembly subsection
%--------------------------------------------------------------------------------------------------
% \subsection{Repeated assembly from a common species pool}
We now attempt to understand the reproducibility of species composition in similar ecosystems. Given many stochastically assembled ecosystems, we can assign each species a `prevalence', i.e. the frequency with which it is observed in one such ecosystem. We want to understand what determines this prevalence.

%--------------------------------------------------------------------------------------------------
% Figure 3
%--------------------------------------------------------------------------------------------------
\begin{figure}
% figure 3
\includegraphics[width=0.45\textwidth]{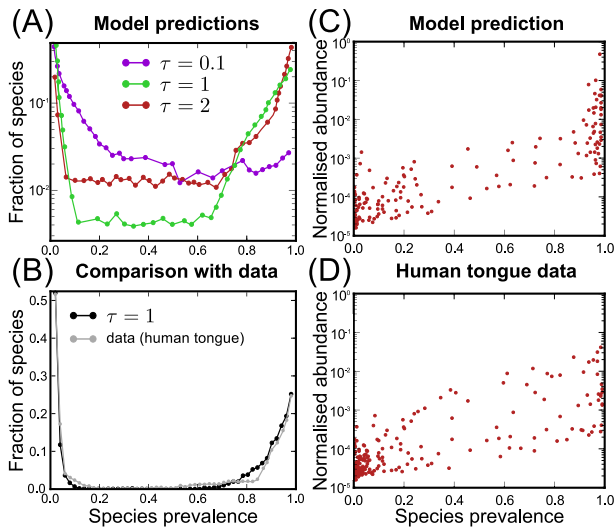}

\caption{\scriptsize{\textbf{Reproducibility from repeated assembly.} (A) Species prevalence distributions from several ecosystems stochastically assembled from a common pool of $1,000$ species (here $\alpha = 0.1$). Shown are distributions for different times $\tau$ in the assembly process (measured in number of immigration attempts by any one species) : 0.1 (violet) at which most species have low-prevalence, 1 (green) and 2 (red) for which we observe a U-shaped distribution (some core species, most peripheral). (B) The distribution for intermediate $\tau$ (black) matches largely with that in longitudinally sampled human oral microbiome (tongue) samples from \cite{Caporaso2011} (gray). (C, D) Normalized species abundance data correlates positively with species abundance in both (C) simulations and (D) oral samples.}}
\label{keyfig3}
\end{figure}

For this, we first generate a `species pool': we run one instance of ecosystem assembly as described above, where over time new species immigrate and attempt to colonize the ecosystem. We add each (even transiently) successful colonizer to our species pool, and continue till our pool has $1,000$ species.  where each species utilizes one resource at a rate $\lambda$ (picked randomly as described above) and secretes two metabolic byproducts as a result of incomplete resource-to-biomass conversion.

With this pool, we simulate several instances of stochastic ecosystem assembly. For each instance, we repeat the aforementioned assembly dynamics under identical initial conditions (i.e. we start with no microbes and just one resource $0$ supplied at flux $\phi_0$). Species attempt to colonize each ecosystem from the pool randomly (with replacement), with the assumption that each species has the same average immigration rate.

The assembly process runs for a fixed time period $\tau$. $\tau = 1$ corresponds to every species attempted immigration \emph{just once} over the assembly process on average, whereas $\tau = 2$ corresponds to every species introduced twice. We repeat our simulations for $\tau = \{\frac{1}{10}, 1, 2\}$.

After collecting several ecosystems for each $\tau$, we plot the distribution of species prevalence --- the fraction of randomly assembled ecosystems in which a species is present (see figure \ref{keyfig3}(A)). We observe that the shape of this distribution depends on $\tau$, as follows:

For small values of $\tau$ (in figure \ref{keyfig3}(A), we show $\tau=0.1$ in violet), the distribution is dominated by small prevalence values. This indicates that --- for this time period $\tau$ --- no species are core, i.e. none of them are found in almost all ecosystem instances. Instead, we find that each species is found at most in a small fraction of them. This makes sense, since at low $\tau$, the stochastic species colonization dominates, and there is hardly any time for species interactions to stabilize parts of the assembling ecosystem.

For higher values of $\tau$ (in figure \ref{keyfig3}(A), we show $\tau=1$ in green and $\tau=2$ in red), the distribution is `U-shaped', i.e. most species are either core (found in most ecosystem instances) or peripheral (found in a small fraction of them). This can be explained as follows: as assembly proceeds, 
species from the pool which use high-flux resources with high resource affinity
end up surviving 
with a high probability. These form the high-prevalence portion of the U-shape. Next, after some delay many other species from the pool that depend on these species at the top trophic layer can successfully colonize these ecosystems. However, stochastic colonization continues to dominate in lowest trophic layers and contributes to the low-prevalence portion of the U-shape. Interestingly, such a U-shaped prevalence distribution is also observed in many real microbial communities: such as longitudinal samples of the human oral microbiome \cite{Caporaso2011} (see the gray distribution in figure \ref{keyfig3}(B); the model prediction is shown in black for comparison) and anaerobic digesters in wastewater treatment plants \cite{Mei2016}.

Moreover, both real ecosystems are characterized by a positive correlation between the prevalence of a species among different samples and their relative abundance in those samples, i.e. species that were found often tend to also dominate their respective communities in abundance (see the scatter plot in figure \ref{keyfig3}(D) for oral microbiome and figure 2(b) in \cite{Mei2016} for wastewater plants). This observation is also captured well by our model: high-prevalence species tend to be be consumers of resources with high effective flux. Since this flux determines steady-state species abundances, our model reproduces this behavior over several orders of magnitude (see model prediction in figure \ref{keyfig3}(C)).

%--------------------------------------------------------------------------------------------------
% Discussion section
%--------------------------------------------------------------------------------------------------
To summarize, high species diversity is routinely observed in microbial ecosystems and the reasons behind this remain a puzzle to microbiologists, ecologists and evolutionary biologists alike \cite{Lozupone2012, Curtis2002, Mei2016}. Central to this puzzle is the competitive exclusion principle \cite{Hsu1977}, which specifies an upper bound on the number of species that can coexist on a given number of resources. For microbial ecosystems in particular, past studies have suggested that facilitation via metabolic byproducts may help alleviate such bounds \cite{Schink1997}. In this study, we present a conceptual model of a microbial ecosystem which demonstrates some ecological consequences of metabolic facilitation --- all of which are borne well by data from real ecosystems. 

In our model, every microbial species is a consumer of a single resource, which --- when utilized for growth --- results in the formation of several metabolic byproducts which may themselves be used as resources by other species. Initially, we supply the ecosystem with one external resource at a constant flux. 
Over time, new species stochastically immigrate and attempt to colonize this ecosystem, their success being determined by competitive exclusion \cite{Hsu1977}.
Both microbes and resources in our model are diluted at a fixed  rate $\delta$. 

Despite its simplicity, this model can generate surprisingly rich ecological behaviors. Specifically, we find that we can tune ecosystem diversity merely by controlling $\delta$, that we can reproduce species abundance distributions for real microbial communities sampled from the human oral microbiome and methanogenic bioreactors, and through repeated assembly, we can capture both the U-shaped prevalence distributions and a positive abundance-prevalence correlation regularly observed in these ecosystems. 

What distinguishes our model from previous `consumer-resource' \cite{MacArthur1970,Pfeiffer2001, Rodriguez2008} approaches? Firstly, we explicitly model energy conservation in the form of incomplete resource-to-biomass conversion and generate metabolic byproducts from what remains. Secondly, we can explicitly handle species abundances to explain why they scale according to a power-law. Finally, we can generate and explain species' prevalence distributions from many microbial communities by simulating several stochastic assemblies. 

Data from microbial ecosystems in different environments corroborate the overarching predictions of this model, namely: human oral microbiome samples \cite{Caporaso2011}, soil communities \cite{Barberan2012}, wastewater treatment plants \cite{Mei2016}, and methanogenic bioreactors \cite{Nobu2015}. Interestingly, the oral data we use is from the human tongue, which is believed to be assembled in a specific temporal order, i.e. late-colonizing species depend on the ones that came before them \cite{Levy2013}. This is very similar to the mechanism behind our simulations.

Note that in the model here we make the assumption that each microbe can use only one resource. In reality, microbes can typically use multiple resources for growth. However, extending our model to allow each microbe to consume more than one resource involves several choices. First, one needs to decide if a microbe would consume resources in parallel or sequentially (both cases are observed in real microbes). Second, one may envisage trade-offs between resource affinity for any single resource and the number of resources. One extreme limit of this trade-off in which the sum of affinities always adds up to the same number has been recently modeled in \cite{Posfai2017}. 
Finally, one needs to distinguish between different `kinds' of resources, e.g. sources of C, N, P, S, Ca, essential vitamins, etc. \cite{Tilman1980, Tilman1982}. In this case, microbes can function as as logical \texttt{AND} gates, i.e. they would have to monopolize a resource of each kind to survive. Moreover, given several resources of each kind, they could also function as \texttt{OR} gates, i.e. choosing any one of each kind. This choice would imply microbes as Boolean logic expressions, their survival being determined by whether the expression encoded in their genome is satisfied by their environment.

Comparing all these options remains beyond the scope of the present study. However, we are currently considering alternative models which incorporate all of these choices. 

\emph{Acknowledgements.} AG acknowledges support from the Simons Foundation as well as the Infosys Foundation.

\bibliography{main}

\clearpage
\onecolumngrid
\renewcommand{\theequation}{S\arabic{equation}}
\setcounter{equation}{0}
\renewcommand{\thefigure}{S\arabic{figure}}
\setcounter{figure}{0}
\section{Supplemental Material}
\section{Derivation for species abundance distribution}
Consider a tree-like ecosystem in a bioreactor as in our model. At steady state, species abundances depend on their effective fluxes corresponding to their consumed resources as in equation (4) in the main text. Consider also the layered arrangement of the ecosystem. Given this, for a species at layer $\ell$ in the tree, its steady-state abundance $b$ is approximately given by $\frac{\langle \widetilde{\phi_{i}}\rangle_\ell (1 - \alpha)}{\delta}$. In the limit of low dilution $\delta$, we can write $b$ roughly as:

\begin{equation}
\label{Beq}
\begin{split}
	\text{log}b &\sim \text{log}\phi_0 - \ell \cdot ( \lvert\text{log}\alpha\lvert + \text{log}(\alpha + \beta) ) + \text{log}(1 - \alpha) + \text{log}\delta \\
    &\sim \kappa - ( \lvert\text{log}\alpha\lvert + \text{log}(\alpha + \beta) ) \cdot \ell,
\end{split}
\end{equation}
where $\kappa$ is a constant. Now, note that that each layer can accommodate $\beta^\ell$ species, where $\beta$ is the number of byproducts per species. To first order, when we ask for the number of species $\mathcal{N}(B > b)$ with abundance greater than a certain value $b$, we are asking for species at layer numbers lower than $\ell_\text{max}$. Inverting equation (\ref{Beq}), we can write this number as follows:

\begin{equation}
\begin{split}
	\mathcal{N}(B > b) &\sim 1 + \beta + \beta^2 + ... + \beta^{\ell_\text{max}} \sim \frac{\beta^{\ell_{\text{max}} + 1}}{\beta - 1} \\
	 				&= e^{\text{log}\beta \big[ \frac{\kappa - \text{b}}{\lvert\text{log}\alpha\lvert + \text{log}(\alpha + \beta)}\big]} \\
                    &= \kappa' \cdot b^{-\big[\frac{\text{log}\beta}{\lvert\text{log}\alpha\lvert + \text{log}(\alpha + \beta)}\big]},
\end{split}                    
\end{equation}
where $\kappa'$ is another constant independent of $b$. From this cumulative distribution, the normalized species abundance distribution will thus be:

\begin{equation}
	\mathcal{N}(B = b) \sim b^{- \big( 1 + \frac{\text{log}\beta}{\lvert\text{log}\alpha\lvert + \text{log}(\alpha + \beta)} \big) }
\end{equation}

\section{Derivation for ecosystem capacity}
We wish to derive the number of species that the ecosystem can accommodate at steady state given a constant dilution rate $\delta$. Note that species cannot survive at steady state unless their steady state abundance is positive. For this to be the case, at the bottom-most layer in the ecosystem $\ell_{\text{max}}$, the resource flux for any consumer species with resource affinity $\lambda$ must be positive. Using the expression in equation (5) of the main text, this implies that the following relation must hold at $\ell_{\text{max}}$:

\begin{equation}
\begin{split}
	\phi_0 \Big( \frac{\alpha}{\beta} \Big)^\ell_{\text{max}} &= \frac{\delta^2}{\lambda} \Big( 1 + \frac{\alpha}{\beta} \Big)^{\ell_\text{max}} \\
    \implies \ell_{\text{max}} &= \frac{\text{log}\phi_0 + \text{log}\lambda - 2 \text{log}\delta}{ \lvert\text{log}\alpha\lvert + \text{log}(\alpha + \beta) }.
\end{split}
\end{equation}

Now, the number of species in the ecosystem $\mathcal{N}_{\text{max}}$ of the order of $\frac{\beta^{\ell_{\text{max}} + 1}}{\beta - 1} = \beta^{\ell_\text{max}} \cdot \frac{\beta}{\beta - 1}$:

\begin{equation}
\begin{split}
	\mathcal{N}_{\text{max}}(\delta) &\sim e^{\text{log}\beta \cdot \big( \frac{\text{log}\phi_0 + \text{log}\lambda - 2 \text{log}\delta}{ \lvert\text{log}\alpha\lvert + \text{log}(\alpha + \beta) } \big) }\\
    &\sim \delta^{-\big(\frac{2\text{log}\beta}{\lvert\text{log}\alpha\lvert + \text{log}(\alpha + \beta) } \big)}.
\end{split}
\end{equation}

\section{Rarefaction curves}
Typically surveys of microbial ecosystems also involve measuring `rarefaction curves', i.e. the number of species observed or detected over the process of sampling several similar ecosystems. Since in our model we performed repeated stochastic assemblies of ecosystems, we can also demonstrate rarefaction curves similar to those observed in the aforementioned surveys. We show below an example of these curves from our model ecosystems in both linear-linear (left) and log-log (right) forms.

\begin{figure*}[h]
\includegraphics[width=0.9\textwidth]{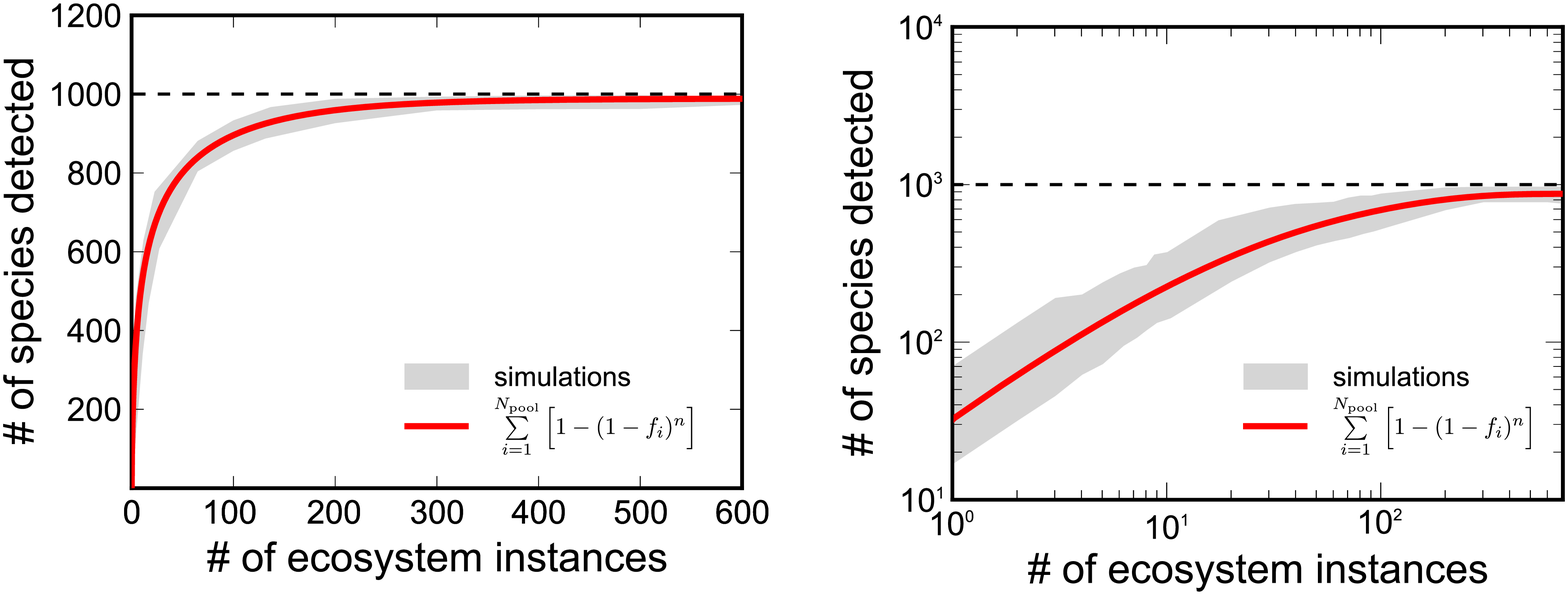}
\caption{The rarefaction curve, that is to say, the total number, $N_{\text{obs}}(n)$, of distinct species observed plotted as a function of the number of samples $n$. Left and right panels show it in linear and logarithmic coordinates respectively. The gray area reflects the variability with respect to the order in which ecosystems are sampled. The red line is the prediction of the Eq. (S6) based on empirically observed prevalences $f_i$ of individual species in the pool. The dashed line is set to $N_{\text{pool}}=1000$, which is the upper bound of $N_{\text{obs}}(n)$.}
\label{suppfig1}
\end{figure*}

We can show that the species' prevalence distributions we discuss in the main text are related to these rarefaction curves. Consider a species pool with $N_{\text{pool}}$ species, each species $i$ with an associated prevalence $f_i$. 

Here $f_i$ represents the frequency with which this species is found in a particular ecosystem sample. We wish to derive $N_{\text{obs}}(n)$, the number of species observed or detected after taking $n$ ecosystem samples.

The probability that after $n$ sampling events, a particular species has \emph{not} been detected is $( 1 - f_i )^n$. Hence, the chance that it is detected at the $n$th sampling event is $1 - ( 1 - f_i )^n$. Summing over all species in the pool, we get the desired expression for $N_{\text{obs}}(n)$:

\begin{equation}
N_{\text{obs}}(n) = \sum\limits_{i=1}^{N_{\text{pool}}} \Big[ 1 - ( 1 - f_i )^n \Big].
\end{equation}

This expression matches our simulated rarefaction curves quite well (the red solid line indicates the expression using species prevalences and the gray envelope indicates results from several simulated ecosystem samples). In our simulations, $N_{\text{pool}} = 1,000$.

Additionally, note that in case the prevalences $f_i$s are sufficiently dense, we may consider only the prevalence distribution $\mathcal{P}(f)$ instead of this discrete sum. In this case, we get the following integral expression over species prevalences $f$:
\begin{equation}
N_{\text{obs}}(n)= N_\text{pool} \cdot \int_{0}^{1} \mathcal{P}(f) \cdot(1- e^{-nf}) df. 
\end{equation}

A practical way to compute it from a known 
prevalence distribution is to first take a derivative of $N_{\text{obs}}(n)$ with respect 
to $n$ given by: 

\begin{equation}
\label{scalingNn}
\frac{dN_{\text{obs}}(n)}{dn} = N_\text{pool} \cdot \int_{0}^{1}f \cdot \mathcal{P}(f) \cdot e^{-nf} df,
\end{equation}
and then integrate the result over $n$. 
Note that $\frac{dN_{\text{obs}}(n)}{dn} \simeq 
N_{\text{obs}}(n+1)-N_{\text{obs}}(n)$; in other words, the number of new 
species detected when the 
number of samples is increased 
from $n$ to $n+1$. 
Hence, it stands to reason that it should 
systematically decrease with $n$ as equation (\ref{scalingNn}) suggests. 

\end{document}